\documentclass{elsart}

\usepackage{graphicx,epsfig,epsf,amssymb}
\usepackage{amssymb}

\newcommand{\vhe}{V\textsc{HE}}

\begin{document}

\begin{frontmatter}

\title{Precision Measurement of Optical Pulsation using a Cherenkov Telescope}
 
\author[HD,LSW]{J. A. Hinton\corauthref{add1}}
\author[HD]{, G. Hermann}
\author[HD,UK]{, P. Kr\"otz}
\author[HD]{, S. Funk}

\address[HD]{Max-Planck-Institut f\"ur Kernphysik, P.O.~Box~103980,
  D-69029~Heidelberg, Germany}
\address[LSW]{Landessternwarte, K\"onigstuhl, D-69117 Heidelberg, Germany}
\address[UK]{now at: Physikalisches Institut, Universit\"at zu K\"oln,
  Germany}

\corauth[add1]{Corresponding author, Jim.Hinton@mpi-hd.mpg.de}

\begin{abstract}
  
  During 2003, a camera designed to measure the optical pulsations of
  pulsars was installed on a telescope of the H.E.S.S. array. The
  array is designed for $\gamma$-ray astronomy in the $\sim$ 100 GeV -
  100~TeV energy regime. The aims of this exercise were two-fold: to
  prove the pulsar timing capabilities of H.E.S.S. on all relevant
  time-scales, and to explore the possibility of performing 
  sensitive optical pulsar measurements using the $\sim\,100$ m$^{2}$ mirror
  of a Cherenkov telescope. Measurements of the Crab pulsar with this
  instrument demonstrate an order of magnitude sensitivity improvement
  over previous attempts using Cherenkov telescopes. 
  Here we describe the design and performance
  of the system and discuss design considerations for future
  instruments of this type.
  
\end{abstract}


\end{frontmatter}

\section{Introduction}
\label{intro}

The six nearby  high spin-down luminosity  pulsars known at
$\gamma$-ray energies provide a laboratory for the study
of high energy processes under extreme conditions~\cite{EGRET_PULSARS}. 
The extremely strong gravitational and magnetic fields close 
to pulsars make the modelling of such objects rather complex.
Indeed, the acceleration processes at work in pulsar magnetospheres, and the 
associated broad-band non-thermal \emph{pulsed} emission, remain
poorly understood despite several decades of effort. Observations in
a wide range of wavelength bands are required to provide information 
not just on different particle energies, but also on different regions within the
magnetosphere. 
Observations above 10~GeV are widely seen as key
to our understanding of the origin of the high energy emission (see
for example \cite{POLARCAP,OUTERGAP}). A small number of 
(apparently pulsed) photons above this energy were detected by
the EGRET satellite but statistics were insufficient to extend
pulsed spectra much beyond 10~GeV~\cite{EGRET_VHE}. The much larger
collection area of ground based instruments is required to explore
the very high energy (\vhe) regime.
The major new air-Cherenkov experiments for ground based $\gamma$-ray
astronomy, such as H.E.S.S.~\cite{HESSproject}, MAGIC~\cite{MAGIC} and
VERITAS~\cite{VERITAS} exhibit greatly improved sensitivity around 100
GeV compared to previous instruments~\cite{STACEE,CELESTE}.
Detections or upper limits on pulsed emission from these new
instruments will likely be based on data taken over months or even
years. Especially in the case of non-detection of a pulsed signal, it
is necessary to prove the long term stability and accuracy of the
timing systems of the experiment.  The key components of
the timing system are the timing hardware (usually based on GPS clock
technology), with associated interface electronics, 
and the software used to derive phase information for an
individual object. Some of these $\gamma$-ray pulsars exhibit
pulsed emission detectable at optical wavelengths. 
The detection of pulsed optical emission using the 
timing system of a $\gamma$-ray instrument can be used 
as a means to demonstrate its absolute timing capability.

Optical pulsar measurements are also interesting in their own right
(see for example \cite{WHYOP}). The six known optical pulsars exhibit
dominantly non-thermal (optical) emission and are all relatively
young, and energetic or nearby. The class of pulsars with measured
optical pulsations therefore falls close to the class of potential
very-high-energy $\gamma$-ray emitting pulsars. The optical emission
also provides a bridge between the well studied radio and X-ray
bands. The recent claim of enhanced optical emission associated with
giant radio pulses (GRPs) of the Crab pulsar~\cite{SHEARER}, acts as
an additional motivation for further studies.

Optical observations of pulsars using conventional optical
telescopes are performed either with fast readout CCD
cameras~\cite{ULTRACAM} or photon counting instruments~\cite{OPTIMA}.
The requirements for such a device are time-resolution 
of $\sim$100 $\mu$s and sufficient signal/noise
to resolve the pulsed emission against the night sky background
(NSB). 
Sensitivity improves linearly with both optical PSF width and
mirror diameter. Traditional devices have excellent angular
resolution and modest mirror area. An alternate concept is to use 
the large mirror area of Cherenkov telescopes to compensate 
for their modest angular resolution. The potential advantages of this
approach include smaller fluctuations
due to \emph{seeing} (intensity fluctuations on small angular scales
caused be atmospheric turbulence), potentially deadtime free
operation and in general a different set of systematic effects.
The idea of measuring the optical pulsations of pulsars with Cherenkov
telescopes is at least a decade old and pulses from the Crab pulsar
have been measured successfully by several instruments:
Whipple~\cite{WHIPPLECRAB}, STACEE~\cite{STACEE_OP},
CELESTE~\cite{CELESTE}, and HEGRA~\cite{CT1CRAB},
with a typical sensitivity of $0.1\,\sigma/\sqrt{t/\mathrm{s}}$ 
(i.e. a one second observation leads on average to an 0.1 standard
deviation detection significance, increasing with the square-root of
time, $t$).
The most recent measurement, using the modestly sized (1.7~m diameter)
mirror of the HEGRA CT-1 telescope, is described in
\cite{CT1CRAB}. This measurement achieved a sensitivity of 0.14
$\sigma/\sqrt{t/\mathrm{s}}$ and the authors estimate a value of 1
$\sigma/\sqrt{t/\mathrm{s}}$ can be achieved for the MAGIC telescope
using the same device (indeed a preliminary study using MAGIC
\cite{MAGIC_OP} confirms this predicted sensitivity. 
However, as all other known optical pulsars are dimmer than the 
Crab by a factor $>200$, even with this level of sensitivity 
over 600~hours of observations would be required to detect pulses from
any pulsar apart from the Crab.

The advantage of the H.E.S.S. telescopes (15~m
focal length, 107~m$^{2}$ of reflector) is that the optical point
spread function is considerably smaller than the pixel size for
on-axis observations, with a FWHM of only
$0.07^{\circ}$\cite{HESSOPT,HESSOPT2}. To exploit this advantage a
single-channel prototype system was installed on the first H.E.S.S.
telescope for 8 nights of Crab pulsar observations in January
2003~\cite{FRANZEN}. A 7-pixel optical pulsar camera with improved
electronic and optical properties was installed in October 2003 on the
last of the H.E.S.S. telescopes to receive a Cherenkov camera.  The
observations described here were performed with this new camera during
October and November 2003.

\section{The Camera System}
\label{camera}

\begin{figure}[ht]
\begin{center}
\mbox{\epsfig{file=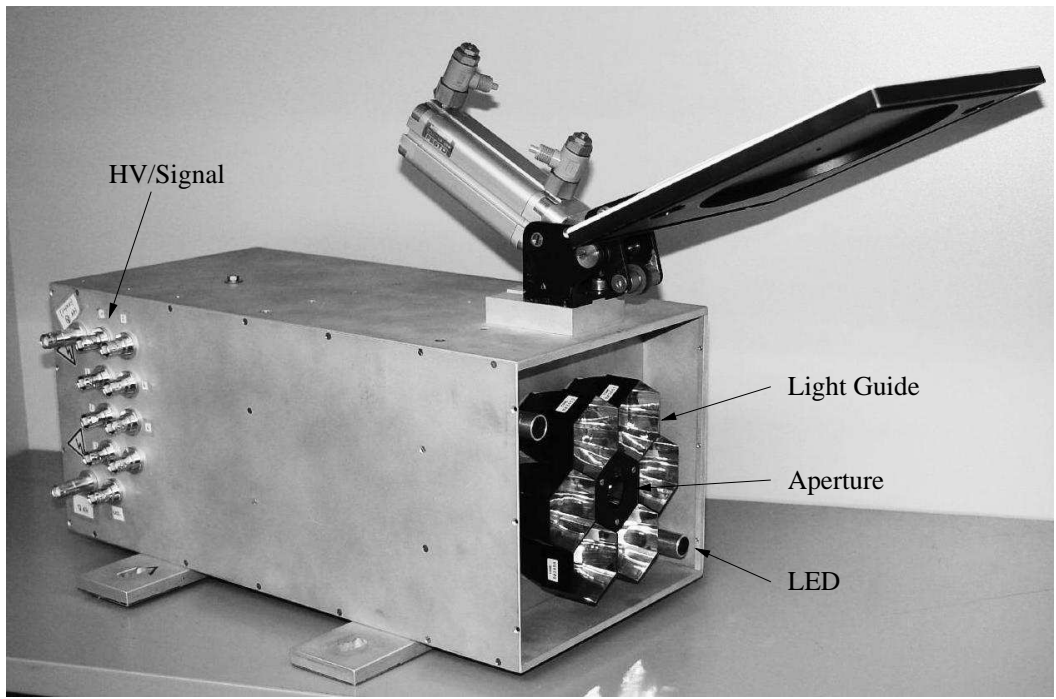, width=1.0\textwidth }}
\end{center}
\caption{Photograph showing the principle mechanical components of the optical
  pulsar camera: the central pixel aperture, the ``veto'' pixel light
  guides, the positioning LEDs and the HV inputs/signal outputs for
  the photomultipliers mounted within the casing.}
\label{fig:Photo}
\end{figure}

\begin{figure}[ht]
\begin{center}
\mbox{\epsfig{file=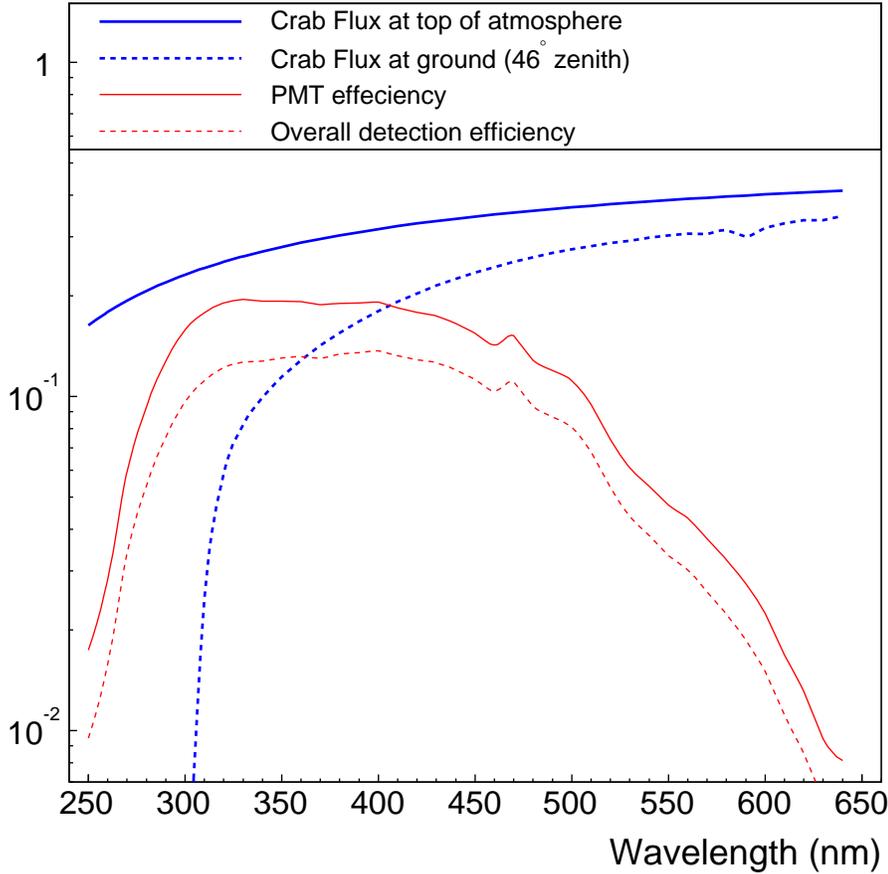, width=0.97\textwidth }}
\end{center}
\caption{Wavelength dependent efficiency of the instrument compared to
the optical spectrum of the Crab nebula~\cite{HUBBLE}, before and
after atmospheric absorption. The lower solid curve gives the combined
quantum and collection efficiency of the PMT. The dashed curve
includes also the reflectivity of the primary mirror and the 
overall collection efficiency of the light cone.}
\label{fig:Wavelength}
\end{figure}

The main improvement over the (single-channel) prototype device described in
\cite{FRANZEN} is the addition of a 6 channel ``veto'' camera surrounding the
central pixel. The additional channels have the dual purpose of
identifying optical atmospheric transients such as air-showers and
meteorites, and of monitoring the NSB level close to the target
pulsar. The optical pulsar camera is shown in figure~\ref{fig:Photo}.
The camera box contains seven photomultiplier tubes (PMTs) with
associated light cones, a high voltage distribution system and two
positioning LEDs.  The Photonis XP2960 PMTs and the light collecting
cones are the same as used for the H.E.S.S. Cherenkov
camera~\cite{HESSCAMERA}. The wavelength dependence of the instrument 
response is given in figure~\ref{fig:Wavelength}. The central pixel is equipped with an
exchangeable aperture (of 1 to 22~mm diameter). 20 or 22~mm apertures
were used for most of the data discussed here, providing optimal
signal to noise given the on-axis optical point-spread-function of a
H.E.S.S. telescope at $46^{\circ}$ zenith angle (the culmination of
the Crab). The 1~mm (pin-hole) aperture is used for pointing tests
using stars.  The camera box is equipped with a pneumatic lid.  The
positioning LEDs are used to monitor pointing and tracking accuracy
via images acquired by a CCD camera mounted at the centre of the
telescope dish, with the pulsar camera in its field of view.

The analogue signals from the seven camera pixels were fed via 50~m
coaxial cables to an electronics container at the base of the
telescope.  The camera electronics were housed in this temperature
controlled environment. Signals from all channels were amplified with
DC-coupled spectroscopic shaper modules, with a characteristic
shaping time of 100~$\mu$s (FWHM). The signal was then digitised with
an over-sampling factor of $\sim\,4$ using a HYTEC VTR 2536 14-bit Flash
ADC. This scheme provides a current measurement proportional to the
average photon flux on 100~$\mu$s time-scales.  Any optical pulsed
signal is measured on top of a DC background generated by NSB photons,
which have a typical rate of $\sim$100~MHz.

Timing information was provided by two Meinberg 167BGT GPS clocks, one
associated with the H.E.S.S. central trigger system~\cite{HESSTRIGGER}
and a second identical clock housed on the telescope tracking room.
These clocks supply 10~MHz and 1~Hz TTL outputs synchronized to
UTC. For the H.E.S.S. clock these signals were distributed via optical
fibre from the central control building.  Two VME counter models are
used to count these TTL pulses and generate event timestamps. The 1~Hz
channel is used to reset the 10~MHz counter. Every two seconds an
ASCII time-string was read via a serial port on the GPS clock, as an
absolute time reference.  The clock counters are read out together
with each FADC sample via a VME based CPU. A custom data acquisition
system (DAQ) developed for this measurement achieved the required
over-sampling factor of $\sim$4 with an average sampling rate 28~kHz
on all channels. Additional electronics in the electronics container
provided temperature monitoring and remote control of photomultiplier
HV and the camera lid.  Data transmission to a remote machine
responsible for data storage occurred in parallel to data taking ---
resulting in deadtime free operation.
Due to the large sampling rate the pulsar camera produced data 
at a rate comparable with that of the H.E.S.S. 960 pixel Cherenkov
cameras ($\sim$1~MB/s).

\section{Observations}

The observations described here were performed in late October and early
November 2003, partly contemporaneous with observations using
the Effelsberg 100~m radio telescope (discussed in \cite{JESSNER}).
The optical measurements described here were accompanied by
$\gamma$-ray observations of the Crab nebula with the 3 remaining
H.E.S.S. telescopes. Overall 60~hours of optical Crab observations
were performed of which 41 hours of on-source observations pass all
data quality criteria. Over 300~GB of raw data were obtained.

The observations consisted of data taken tracking the Crab pulsar
itself, interspersed with occasional off-source runs and periodic
pedestal monitoring runs taken with the camera lid closed.
Observations occurred semi-automatically, with intervention from the
H.E.S.S. shift-crew only at the beginning and end of each night. To
keep the camera centred on the pulsar it was necessary to make online
corrections for atmospheric refraction and bending in the arms of the
telescope.  The verification of the absolute pointing of the
instrument was done by scanning several stars across the aperture of
the central pixel, and also using CCD images of the positioning LEDs
and star images on the camera lid. The telescope tracking precision
was monitored during the measurement and exhibited a maximum rms
deviation of 3~$''$.  The total systematic pointing error is estimated
as $\sim 30''$, introducing a negligible error on the recorded optical
signal.

For roughly half of the observing time, two independent GPS clocks
were read-out in parallel: the H.E.S.S. central trigger timing system
and the independent system of the optical pulsar instrument.  The mean
difference between the two event time-stamps during this time was
2.1~$\mu$s, consistent with the length of the optical fibre over which
the central clock time was transmitted. The rms of this time
difference was 150~ns, consistent with the precision quoted by the
clock manufacturers.

The night sky background level in the camera FOV is rather
inhomogeneous. The Crab nebula has a V-band magnitude of 8.4 and a
diameter of $\approx$ 6$'$, covering essentially the entire central
pixel. The brightest star present in the surrounding pixels has
$M_{\mathrm{v}}$ = 9.9, causing significant variation in pixel
currents as it rotates around the outer pixels.  The mean NSB level in
the outer pixels was found to be of $\sim 3\times10^{12}$ photons
m$^{-2}$ s$^{-1}$ sr$^{-1}$.  Previous measurements of the NSB level
at the H.E.S.S. site found $\approx2.4\times10^{12}$ photons m$^{-2}$
s$^{-1}$ sr$^{-1}$ (300-650~nm) close to the zenith and levels higher
by a factor $\sim\,2$ on the galactic plane in the outer
galaxy~\cite{HESS_NSB}. NSB measurements of this field using a HEGRA
telescope on La Palma~\cite{CT1CRAB} yielded a value of
$4.3\times10^{12}$ photons m$^{-2}$ s$^{-1}$ sr$^{-1}$, in reasonable
agreement, given the different site and the different altitude range
of our observations.

\section{Analysis \& Results}

The first step of the analysis is the summation of every four
consecutive samples to produce statistically independent measurements
of $\approx$ 140\,$\mu$s duration.  A timestamp is generated based on
the average time of the four samples. For each five minute observation
run, pedestal values are subtracted based on the ADC values with the
camera lid closed.  Periods with unstable weather conditions are
removed by cuts on the rms and gradient in the signals measured in the
veto pixels. Transient background events are removed by excluding 2~s
blocks in which the signal in any \emph{surrounding} pixel exceeded 5
times the signal rms for any 5 consecutive measurements. No cuts are
made based on the central pixel signal.

\subsection{Timing and Light-Curve}

A crucial aspect of the analysis is the barycentering and
phase-folding of the event times. For this purpose software developed
specifically for H.E.S.S. was employed~\cite{SGPhD}. The software has
been compared against the standard TEMPO package~\cite{Till,TEMPO} and no
difference greater than 1~$\mu$s found for non-binary pulsars.
Ephemerides from Jodrell Bank were used for the phase
calculation~\cite{Jodrell}.  Comparing the ephemerides provided for
October and November we derive a second derivative of the pulsar
frequency: $\ddot{f}\,=\,
9.3\,\times\,10^{-21}\,\mathrm{s}^{-3}$. This value of $\ddot{f}$,
together with $\dot{f}$, $f$ and $t_{0}$ taken from the published
November 15$^{\mathrm{th}}$ ephemerides, was used to calculate the
absolute phase information given here.

\begin{figure}[t]
\begin{center}
\mbox{\epsfig{file=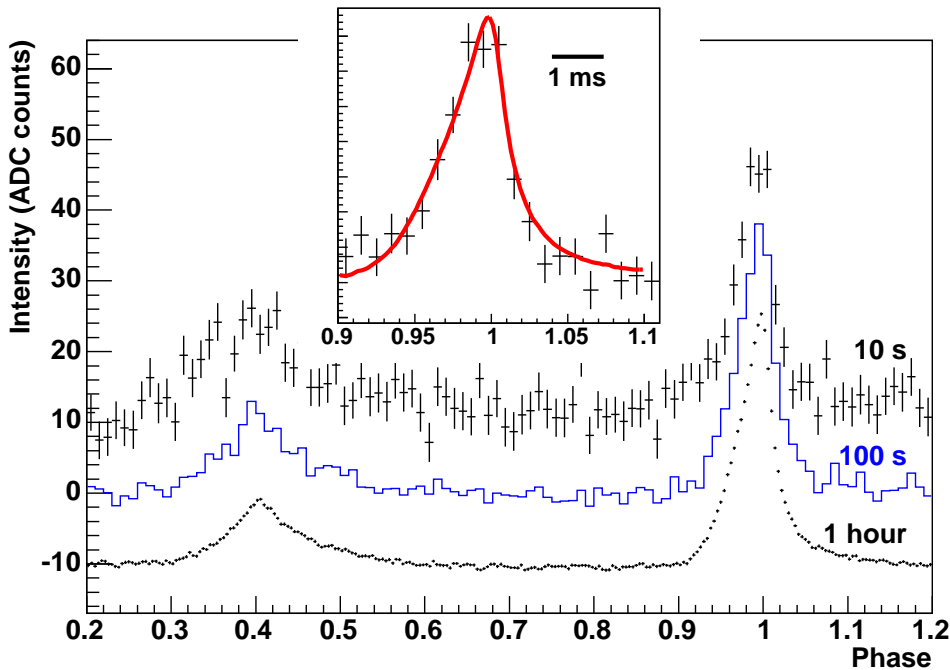, width=1.01\textwidth }}
\end{center}
\caption{Average optical signal versus phase for the Crab pulsar for 10
  second, 100 second and 1 hour integration times. The inset shows 
  a fit to the main peak for the 10 second dataset. For clarity, 
  offsets of $\pm$10 counts have been added to the 10 s and 1 h
  datasets. The DC signal produced by the Crab nebula and the NSB 
  has been subtracted ($\sim 2 \times 10^{4}$ ADC counts).
}
\label{fig:Phaso10sec}
\end{figure}

Following phase-folding a clear signal from the Crab pulsar was
visible in all datasets. The average signal/noise of the full dataset
is such that an average significance of $4\,\sigma/\sqrt{t/\mathrm{s}}$ can be
assigned to the pulsed signal, more than an order of magnitude better
than previous Cherenkov telescope measurements.  The level of
precision achievable on different time-scales is illustrated in
figure~\ref{fig:Phaso10sec}. Phase-folded light-curves extracted from
10\,s, 100\,s and 1\,hour datasets are shown. Even in a 10\,s
exposure, the peak position is resolved (inset in
figure~\ref{fig:Phaso10sec}) to better than 1\,ms.
The position of the peak in subsets of the data is determined 
by a fit of a smoothed version of the overall measured phasogram
with two free parameters: the normalisation and the relative 
phase shift. This empirical fit function is shown in the inset 
of figure~\ref{fig:Phaso10sec}.

\begin{figure}[t]
\begin{center}
\vspace{6mm}
\mbox{\epsfig{file=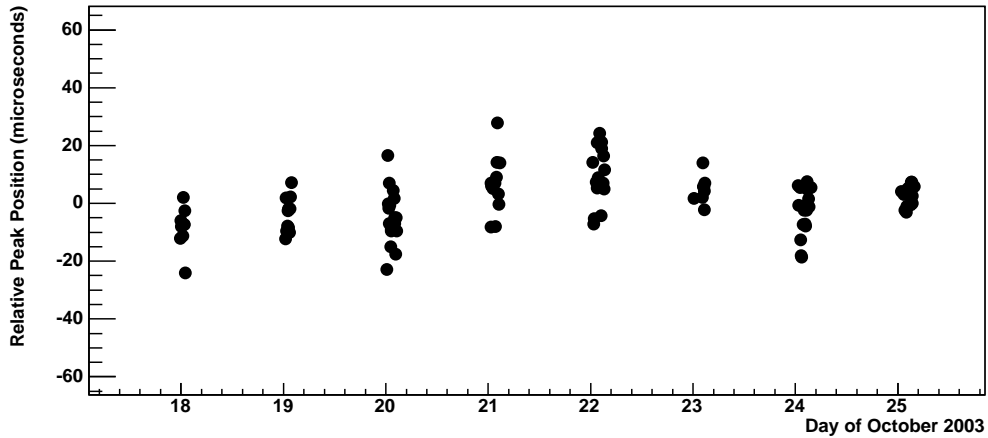, width=1.07\textwidth }}
\end{center}
\caption{Reconstructed relative main peak position for the Crab pulsar over one
  week in October 2003. Each point represents one 5 minute run.}
\label{fig:PeakDrift}
\vspace{3mm}
\end{figure}

As can be seen from figure~\ref{fig:PeakDrift} the position of the
main peak can be measured with an accuracy of $<20\,\mu$s in one
5-minute run. Unfortunately the available radio ephemerides have an
absolute precision of $\sim 70\mu$s, so any comparison is limited by
the radio accuracy.
The apparent drift during the period of our measurements ($\sim$ one
month) was $\approx$60 $\mu$s, consistent within the accuracy of the 
Jodrell ephemerides for October and November. The apparent phase shift
of the mean peak between January and November 2003 was 
$\approx$100 $\mu$s, again consistent within the accuracy of the 
radio ephemerides used.

\begin{figure}[h]
\begin{center}
\mbox{\epsfig{file=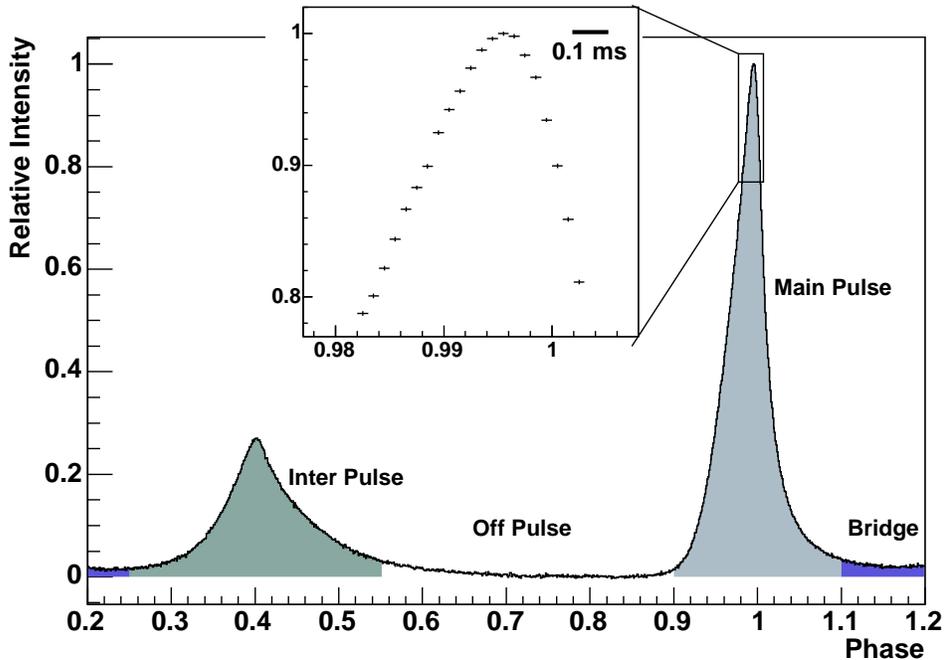, width=1.0\textwidth }}
\end{center}
\caption{ Phasogram extracted from the complete 39 hour dataset. 
  The main phase regions are marked: the bridge emission (0.1-0.25), 
  the inter-pulse (0.25-0.55), the main pulse (0.9-1.1) and the off
  pulse region (0.55-0.9). The inset shows the 3\% of
  the phase around the main peak position on an expanded scale.
  The 100~$\mu$s scale bar in the inset corresponds to the shaping
  time of the measurement, any structures in the light-curve on 
  shorter timescales are unresolvable.
}
\label{fig:Phasogram}
\end{figure}

The phasogram extracted from the full 39~hour dataset is shown in
figure~\ref{fig:Phasogram}. The best published optical phasogram for
the Crab nebula is that extracted from 2~hours of observations with
the Hubble Space Telescope (HST)~\cite{HUBBLE}. 
We find generally good agreement between the H.E.S.S. result and that
of HST. For example, the ratio of the height of the main pulse to that
of the inter-pulse: HST 3.78$\pm$0.11, H.E.S.S. 3.721$\pm$0.003; and
the FWHM of the main peak: HST 0.0431$\pm$0.0003,
H.E.S.S. 0.0445$\pm$0.0001.  The H.E.S.S. measurement is 40 $\mu$s
wider than that from HST, consistent with the expected smearing
introduced by imperfect ephemerides over the long exposure
time. Indeed, fits to all 489 individual 5 minute phasograms show a
mean value of 0.0432$\pm$0.0001, identical to the HST result.  In our
data the phase position of the main optical pulse precedes that of the
radio peak by 134$\pm$2 $\mu$s, again consistent with previous
measurements (100~$\mu$s \cite{SHEARER}) within the error introduced
by the imperfect radio ephemeris.

The signal measured with our instrument corresponds approximately to the
wavelength range 300-650~nm, with a rather non-uniform response
illustrated in figure~\ref{fig:Wavelength} (more details on the individual optical
components can be found in~\cite{HESSOPT}).
The most meaningful way to compare the pulsed flux measured here with 
previous measurements is to convolve the measured Crab pulsed spectrum
with our wavelength dependent response function. Using the spectrum
reconstructed by~\cite{HUBBLE} we predict a pulsed signal of 1000
photoelectrons per pulse (at 46$^{\circ}$ zenith). 
The mean measured value is 1200 p.e./pulse, in good agreement within 
the $\approx$30\% systematic error introduced by uncertainties in 
instrumental reflectivities, efficiencies and atmospheric transmission.

\subsection{Search for Giant Pulses}

Extreme pulse height variability is a common feature of radio pulsars.
Giant radio pulses (GRPs) are normally coincident with main pulse and inter-pulse ~\cite{SHEARER}
but can apparently also occur in other phase
regions~\cite{JESSNER}. While an analogous phenomenon has not so far been 
seen in the optical, an average 3\% enhancement in flux of the optical
pulse, in coincidence with GRPs, has been reported~\cite{SHEARER}.
Given the signal/noise of our measurement a few $\times 10^{4}$
simultaneous GRPs would be required to confirm this value. 
However, given the relatively long duration of
our measurement it is useful to derive a limit on the rate of 
\emph{large} optical pulses, independent of the radio pulse height.

\begin{figure}[ht]
\begin{center}
\mbox{\epsfig{file=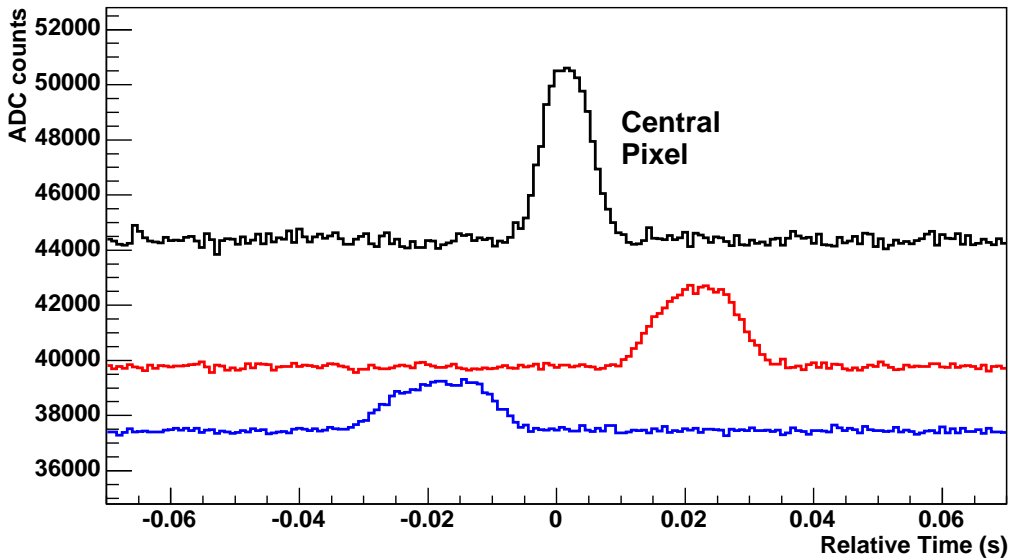, width=1.0\textwidth }}
\end{center}
\caption{Light-curve in three pixels indicating the passage of a
  meteorite candidate. The central pixel (upper curve) shows a narrower 
  profile than that of the outer pixels (lower curves) due to its reduced aperture.}
\label{fig:Meteorite}
\vspace{3mm}
\end{figure}

\begin{figure}[h]
\begin{center}
\mbox{\epsfig{file=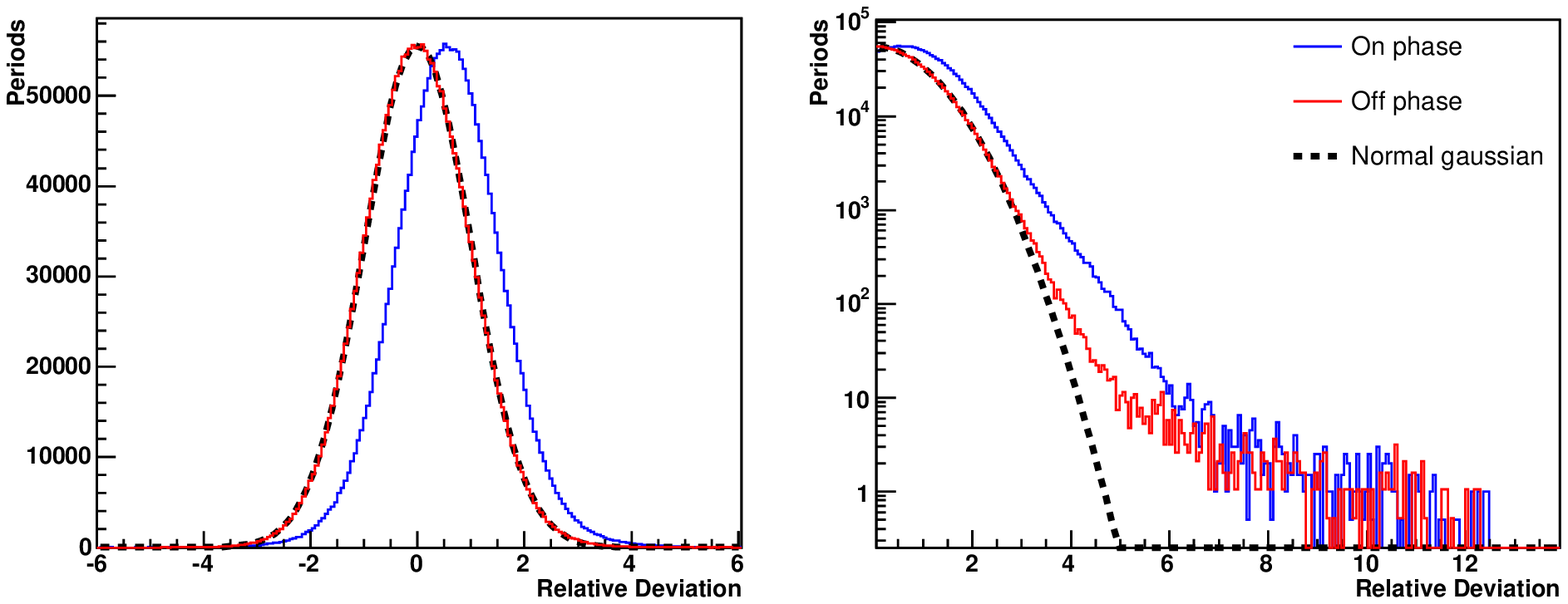, width=1.01\textwidth }}
\end{center}
\caption{Relative fluctuations in the integrated signal in on- and
  off-phase regions for each pulse (linear scale - left and log. scale
  - right). The on-phase region corresponds to the main pulse and 
  inter-pulse regions shown in figure~\ref{fig:Phasogram}. Off-pulse
  here corresponds to a combination of the true off-pulse 
  region and the bridge region as shown in figure~\ref{fig:Phasogram}.
  After tight quality selection to 
  remove optical transients and unstable weather, 2.2 million pulses
  remain. The off-phase distribution follows closely a normal 
  gaussian distribution (dashed-line) as expected, but with a 
  significant tail to large values. This tail can be attributed to
  transient events occurring only in the central pixel. A similar
  tail is seen on the on-phase distribution.
}
\vspace{3mm}
\label{fig:AmplitudeDists}
\end{figure}

The major background for such a search in our dataset is meteorites. These are normally 
detected in several pixels and take $\approx$50~ms to cross the
camera. Such events, as for example that shown in
figure~\ref{fig:Meteorite}, are readily rejected from the analysis
using the outer pixel information.
However, an irreducible background of events travelling almost on-axis 
and hence illuminating only the centre pixel, is present.
Figure~\ref{fig:AmplitudeDists} shows the distribution of integrated 
signal in each pulse (or on- and off- phase regions) normalised by the 
expected poisson fluctuations.
No fluctuations $>$ 20$\times$ the mean pulse amplitude were observed
during our measurement, neither on- nor off-phase. We can therefore
place a limit on the fraction of such large pulses of $1.4 \times\,10^{-6}$
(95\% confidence). 
This result should be compared to the situation at radio frequencies 
where pulses exceeding the mean flux by several orders of magnitude 
are frequent. 
We note that our instrument is sensitive to 
giant pulses of very short ($<$ 1 $\mu$s) duration, as seen in
the radio band, which may not be the case for photon counting 
instruments. 


\section{Summary and Outlook}

We have constructed and tested an optical pulsar monitoring system for
installation on Cherenkov telescopes.  We have shown that such a
system can approach the sensitivity of conventional optical telescopes
(with custom built cameras) in measuring short-time-scale
(millisecond-second) optical pulsations. In addition, the excellent
agreement on the shape of the Crab pulsar light-curve validates the
timing hardware and software used for H.E.S.S., thus demonstrating the
validity of pulsed emission limits from
H.E.S.S~\cite{pulsedlimits}. We also derive an upper limit on the
frequency of \emph{giant pulses} which is complementary to existing
limits and measurements in that it includes also very short ($\sim$
nanosecond) pulses.

As the energy threshold of the IACT technique are pushed down further
(e.g. with the second phases of both the H.E.S.S. and MAGIC
experiments), the detection of pulsed VHE $\gamma$-ray emission will
become more likely, thereby increasing the importance of optical monitoring
devices.  We consider two possibilities for future instruments of this
type.  Firstly a small device for monitoring purposes, gathering data
in parallel to $\gamma$-ray with the Cherenkov camera. Ideally such an
instrument would replace one pixel of the Cherenkov camera.  The
problem with this approach is the observing strategy of modern
instruments, for example 'wobble' mode and convergent pointing, both
of which move the target source away from the centre of the field of
view of the telescopes.  Field rotation then prevents the observation
of a source in a signal pixel through an entire run.  The second
possibility is to adapt the photo-sensor current monitoring of a
Cherenkov camera to provide high rate ($>$1 kHz) and resolution
sampling of the sky brightness. With integration and sampling rates on
comparable time scales such a camera could be used to measure optical
pulsations (and transients) anywhere in the $\sim\,4^{\circ}$ field of view of the
instrument, in parallel to $\gamma$-ray observations. A camera with these properties is currently under
test~\cite{SPC}.  For the 600~m$^{2}$ class telescopes of the
next generation, 
deep observations of young pulsars with such a system could lead to
the discovery of new optical counterparts to known radio pulsars. Such
a system is also desirable in the search for MeV/GeV emission from
short time-scale GRBs~\cite{SGARFACE}.

\section*{Acknowledgements}

The authors would like to acknowledge the support of their host
institutions, and additionally support from the German Ministry for
Education and Research (BMBF).  We appreciate the excellent work of the engineering and
technical support staff in Heidelberg and Namibia in the construction and operation of the
equipment. We would like to thank the entire H.E.S.S. collaboration 
for their cooperation and assistance, in particular: F.\ Breitling,
K.\ Bernl\"ohr, O.\ Bolz,  S.\ Gillessen, M.\ Holleran.

\end{document}